\documentclass{PoS}

\title{Anomalies in the GRB spatial distribution}

\ShortTitle{Anomalies in the GRB spatial distribution}

\author{\speaker{Istvan Horvath}\\
        National University of Public Service, Budapest, Hungary\\
        E-mail: \email{horvath.istvan@uni-nke.hu}}

\author{Zsolt Bagoly\\
       E\"otv\"os  University, Budapest, Hungary\\
        E-mail: \email{zsolt@elte.hu}}
\author{Jon Hakkila\\
       College of Charleston, Charleston, SC, USA\\
          E-mail: \email{HakkilaJ@cofc.edu}}
\author{L. Viktor T\'oth\\
       E\"otv\"os  University, Budapest, Hungary\\
         E-mail: \email{L.V.Toth@astro.elte.hu}}
\abstract{

Swift's remarkable ability to quickly localize gamma-ray
bursts has led to the accumulation of a sizable
burst sample for which both angular locations and 
redshifts are measured. 
This sample has become large enough that it can
potentially be used to probe angular anisotropies 
indicative of large-scale universal structure. 
In a previous work, a large clustering of gamma-ray 
bursts at redshift $z \approx 2$ 
was reported in the general direction
of the constellations of Hercules and
Corona Borealis. Since that report, a $42\%$ 
increase in the number of $z \approx 2$ gamma-ray bursts 
has been observed, warranting an updated analysis.
Surprisingly, the cluster is more pronounced 
now than it was when it was first reported.

}

\FullConference{Swift: 10 Years of Discovery\\
         2-5 December 2014\\
         La Sapienza University, Rome, Italy}

\begin{document}

\section{Introduction}\label{sec:intro}
Gamma-ray bursts (GRBs) are
the most luminous objects known; they are luminous 
enough that their positions can potentially be
used to help map out
large-scale universal structures.
GRBs are tracers of the stellar matter from which
they formed, and they significantly outshine their host 
galaxies so that they can be mapped even when the 
distribution of underlying galaxies cannot.
The drawback is that the detection rate is small
(95 per year by Swift; \cite{sak11}),
and the rate at which redshifts are measured for the
detected bursts is even smaller (roughly 35 per year).

The angular distribution of GRBs has been
studied in detail over the past two decades
(\cite{bri96,bal98,bal99,mesz00,mgc03,vbh08}).
Initially, the distribution's angular isotropy
was examined in response to the 
hypothesis that GRBs had Galactic origins. 
After the cosmological nature of GRBs was established,
however, the focus of isotropy studies shifted to subsamples
having potentially different angular distributions
(\cite{bal98,cli99,mesz00,li01,mgc03,vbh08}). 
These studies, originally dependent on GRBs
for which redshifts had not been measured,
became more reliable as the number of GRBs 
with known redshifts increased, but remain
limited by small sample sizes.
Swift's compilation of a large number of GRBs
having known redshifts
has reinvigorated large-scale isotropy studies,
although the small numbers of GRBs in specific
redshift ranges (corresponding to radial shells) 
limits these studies to the detection
of large, pronounced anisotropies.  

We have recently identified a surprisingly large anisotropy 
suggestive of clustering in the GRB angular distribution 
at around redshift $z \approx 2$ (\cite{hhb13,hhb14})
in the general directions of the constellations of Hercules
and Corona Borealis. 
The scale on which the clustering occurs is 
disturbingly large: the underlying distribution of
matter suggested by this cluster is big enough to question
standard assumptions about Universal homogeneity
and isotropy. Fortunately, Swift's continued
detection of GRBs makes the hypothesis testable:
if the anisotropy is attributable to statistical sampling,
then the cluster should become less pronounced
as more GRBs are detected.

As of November 2013, the redshifts of 361 GRBs have been 
measured\footnote{http://lyra.berkeley.edu/grbox/grbox.php}
with the sky distribution shown in Fig.1; most of these GRBs were detected
by Swift. This sample represents
a 28\% increase over that used in our
previous analysis (283 bursts observed until July 2012). 
The number of GRBs in the 
$1.6 < z < 2.1$ redshift range (\cite{hhb13})
(where the cluster resides) has increased from 31 bursts
to 44 bursts, a 42\% sample size increase that is large enough
to warrant an updated analysis.

\begin{figure}[h!]\begin{center}
  \resizebox{.95\hsize}{!}{\includegraphics[angle=270]{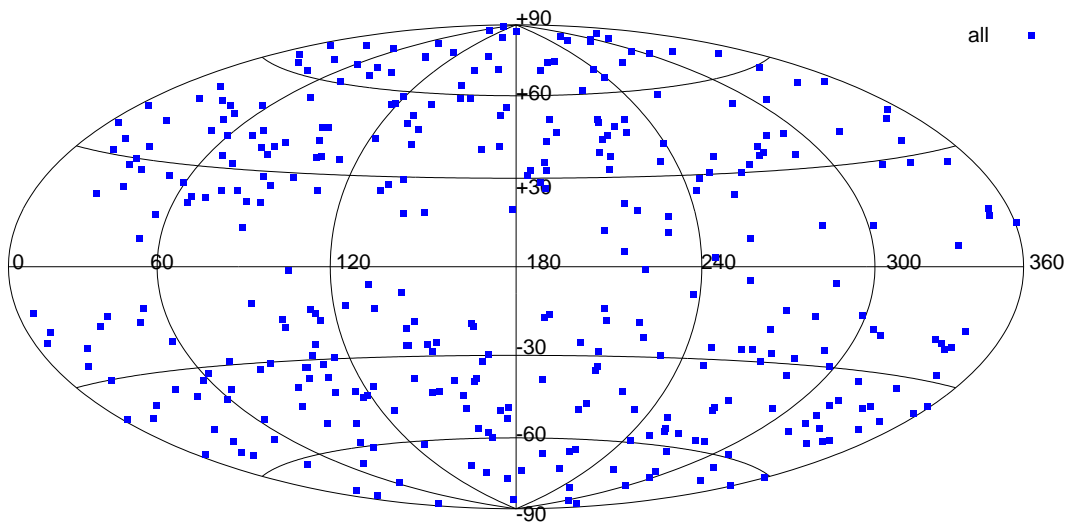}}
  \caption{\small{ The sky distribution of GRBs with measured redshift. Although the distribution of all GRBs is
		fairly isotropic, extinction causes this sample to miss GRBs
		near the Galactic plane.
  }}
  \label{fig:0906}
\end{center}
\end{figure}

\section{Data Analysis}\label{sec:GRB spatial distribution}
   We analyze the angular distributions in each
or our predefined redshift $z$ bins using 
the k$^{\tt th}$ nearest neighbor 
and the bootstrap point radius methods.
These statistical tests are chosen so
that we can directly compare our
new results with our previous ones. 

The redshift (radial) bins used in this study
are chosen to be the same
as those defined in our previous work.
In theory the small uncertainties
associated with redshift measurements
allow the GRB sample to be easily 
subdivided into as many redshift bins 
as desired. In practice, small number statistics
limit the confidence with which 
anisotropies can be detected in each bin.

We have subdivided the total sample of 361 
bursts into different numbers of
redshift bins ranging from two bins
(note that a choice of one bin
corresponds to the bulk angular GRB
distribution in the plane of the sky)
to nine bins.
These choices not only allow us to explore
the angular characteristics of a variety of 
radial bins corresponding to
redshift intervals, but also allow us to 
identify the redshift range within which 
any anisotropies lie, should we
discover them.
The choice of nine radial bins provides us with narrow z-bins 
having the smallest number of bursts per bin ($\approx 40$) for
which we feel we can make reasonable, quantifiable estimates 
on bulk anisotropies. 
When choosing between 2 and 9 radial divisions, we
select the bin sizes that allow us to maintain similar
numbers of bursts in each radial bin. 
When we choose a number of bins that does not
allow an equal number of bursts to be placed
in each bin, then we redefine the bin boundaries
so that the excess GRBs are those with the smallest
redshifts; these are subsequently excluded from the analysis.

\begin{table}[t]\begin{center}
    \hfill{}
    \caption{An example of the $31^{\tt st}$ nearest-neighbor 
    test for four radial groups, with redshift
    boundaries defined in the text. 
    Tabulated numbers represent the 
    KS-test significance that two groups have different
    $31^{\tt st}$ nearest-neighbor distributions.
    Boldface type indicates that significant 
    (more than 3$\sigma$) differences exist
    between group 2 ($1.61 \le z < 2.68$)  
    and other radial groups.}
	\begin{tabular}{|l||c|c|c|c|c|c|c|c|c|}\hline 
		 & $z_{min}$ &  gr2  & gr3 & gr4  \\ \hline \hline 
gr1 & 2.68 &  {\bf 0.9999999}  & 0.942 & 0.672  \\ \hline
gr2 & 1.61 &    & {\bf 0.99904}  & {\bf 0.9999988}   \\ \hline
gr3 & 0.85 &    &  & 0.960   \\ \hline
	\end{tabular}
	\hfill{}
	\label{tab:err}
\end{center}
\end{table}

The first statistical test we apply to the radially binned 
distributions is the k$^{\tt th}$ nearest-neighbor statistic; 
this test looks at the angular separation between each burst
and the k$^{\tt th}$ closest burst to it. When this test considers 
only the nearest neighbor ($k=1$), it is sensitive to anisotropies
on small angular scales corresponding to paired bursts.
When looking at widely-separated bursts with large $k$ values,
the test is sensitive to anisotropies on much larger
angular scales.

We apply the k$^{\tt th}$ nearest-neighbor statistic 
to all burst pairs in each of our radial bins, for the bins
in each sample of two to nine radial bins. Most of
the times the results are statistically consistent with 
isotropy, regardless
of the value of $k$, the number of radial bins chosen, or
the particular radial bin being examined.
The exceptions arise in the radial bin
containing GRBs with redshifts of $z \approx 2$,
where anisotropies are observed for medium-sized
$k$ values consistent with clustering. 
These anisotropies are most pronounced
in the $1.6 \le z < 2.1$ radial bin, to which
this clustering appears confined. 
This result is consistent with our previous findings 
for a smaller data set \cite{hhb13,hhb14}.

We demonstrate typical results based on a choice of 
four radial bins. Each bin contains 90 GRBs, with the 
bins defined by $2.68 \le z < 9.4$ (group 1), 
$1.61 \le z < 2.68$ (group 2), $0.85 \le z < 1.61$ 
(group 3), and $0 \le z < 0.85$ (group 4).
Table 1 shows the significance of the null hypothesis 
for this example using the
Kolmogorov-Smirnov test that the two distributions 
are different. Boldface type indicates that the significance 
that the two group's 31st 
nearest-neighbor distributions differ
by more than 3$\sigma$. There are no significant 
differences within the
group 1, group 3, or group 4 distributions, but the 31st 
nearest-neighbor distributions in group 2 indicate
a significant anisotropy. The same indication of a large-scale
anisotropy in this radial group is found from all nearest
neighbor distributions spanning the range $22 \leq k \leq 55$,
indicating that the anisotropy occurs on an
angular scale of intermediate size.


The significance, angular size, and location of the large, 
loose GRB cluster in the redshift range
$1.6 < z \le 2.1$ can also be estimated using
the bootstrap point-radius method
described in section 5 of Horvath et al. 2014 (\cite{hhb14}).
This test compares the 44 GRBs found in the
$1.6 < z \leq 2.1$ radial bin to 44
randomly selected GRBs drawn
from the rest of the sample.
This technique counts the number 
of $1.6 < z \le 2.1$ bursts within 
a circle of predefined radius 
$\theta$ surrounding
a random 'cluster center' location.
Comparison samples are created by randomly
drawing 44 bursts from the rest of the
dataset and counting the number of 
bursts lying within the same angular circle.
Statistics are generated by repeating
this process 10000 times and 
counting the largest number of bursts
found within the circle during these runs.
Once results have been obtained,
$\theta$ is increased and the process
is repeated for eighty different $\theta$ values 
in equal area steps ranging from
$\theta = 12.84 ^\circ$ to $\theta = 180 ^\circ$.



The Monte-Carlo bootstrap point-radius method
verifies that the anisotropy found in the $1.6 < z \leq 2.1$ 
redshift range is consistent with angular clustering.
Forty nine of the 80 angular clustering scales
tested exhibit excess numbers of bursts within
the defined $\theta$ limits.
For example, using the measurements from the cluster center locations
producing the largest GRB clusters,
an angular circle having
a radius of $\theta = 22.3 ^\circ$ (corresponding
to $3.75\%$ of the sky) is found to contain 13 of the 44 bursts ($30\%$),
a circle with radius $\theta = 34.4 ^\circ$ (corresponding
to $8.75\%$ of the sky) contains 18 of the 44 bursts ($41\%$),
and a circle with radius $\theta = 51.3 ^\circ$ (corresponding
to $18.75\%$ of the sky) contains 25 of the 44 bursts ($57\%$).
Only two of 17500 bootstrap cases had 25 or more GRBs inside this
latter circle indicating a statistically significant (p=0.0001143) deviation 
(the binomial probability for this being random is  $p_b=2\times 10^{-8}$).



The $42\%$ increase in sample size should have
noticeably decreased the significance of the $1.6 < z \leq 2.1$
cluster if random sampling was responsible for its existence.
However, the cluster has become more pronounced in the 
$4^\circ \leq \theta < 90^\circ$ angular
radius range as more GRBs have been added to the sample.


In this range of angular radii, 49 angular circles contain enough GRBs to
exceed the $2\sigma $ significance level (compared to 28 found in our 
previous analysis \cite{hhb14}).
Additionally, there are 16 angular circles containing enough
GRBs to exceed the $3\sigma$ level (compared to only 2 in our
previously published result).
Therefore, the evidence has strengthened that these bursts are 
mapping out some large-scale universal structure.

Both of the aforementioned tests are independent
of sky exposure; by calculating only the
relative angular positions of the detected bursts 
the techniques assume simply that whatever biases
present at different redshifts are the same
as those in the $1.6 < z \leq 2.1$ redshift
range.

\section{Discussion}
Gamma-ray bursts are not distributed
isotropically in the $1.6 \le z < 2.1$ redshift range;
they exhibit evidence of
large-scale clustering. This clustering,
first identified in 2013 (\cite{hhb13,hhb14}), has
become more pronounced with recent
GRB detections by Swift, supporting the
idea that the clustering may be real rather than
due to a statistical variation in the detection rate.
The $k^{\tt th}$ nearest 
neighbor test indicates that
GRBs in this redshift range are
likely to have more neighbors at moderate
angular separations 
than those at other redshifts. The two-dimensional
point-radius method also finds evidence for a
large-scale angular clustering in this redshift
range; the angular diameter encompassed by this 
clustering is likely many tens of degrees across.

The aforementioned techniques can be used to demonstrate
the existence of a large, nebulous GRB cluster, but the nature 
of the tests used here prevent us from knowing exactly 
where the cluster is located, what its structure is like, 
or how big it might be. 
Selection biases
due to instrumental sky exposure and 
visual extinction by dust complicate
this interpretation by reducing the
rate of detection in some parts of the
sky relative to others. Based on the
detected bursts, the cluster 
covers roughly one-eighth of the sky,
and seems to
encompass half of the constellations of Bootes, 
Draco, and Lyra, and
all of the constellations of Hercules and Corona Borealis.
The name of the structure has been popularized as the
{\em Hercules-Corona Borealis Great 
Wall}, or Her-CrB GW. However, we note that
sampling biases could cause
the cluster to be offset by many degrees from where it 
currently appears to be.

Because GRBs are the most luminous, energetic objects known, they
are tracers of the presence of normal matter that can be
detected at distances where the matter is otherwise too 
faint to be observed. Based on the analysis performed
here, we estimate the size of the Her-CrB GW to be about 2000-3000 Mpc
in diameter. Few limits on its radial thickness exist, other than the fact that it
appears to be confined to the $1.6 \le z < 2.1$ redshift range.
This large size makes the structure inconsistent with current
inflationary Universal models, as 
it is larger than the roughly 100 Mpc limit thought to signify
the ``End of Greatness'' at which large-scale structure ceases.

However, the Her-CrB GW is not the first optical/infrared structure found
to exceed the 100 Mpc size limit.
In the 1980s, Geller and Huchra (\cite{gel89}) mapped 
galaxies and galaxy clusters
in a portion of the sky to $z\approx0.03$ and found 
a 200 Mpc size structure that was later called the CfA2 Great Wall. 
In 2005 an object twice this size named 
the Sloan Great Wall (\cite{gott05}) was reported.
In the ensuing years, several other large filamentary structures
have also been identified.
Roger Clowes and his team have found several large
clusters of luminous quasars; the largest of these
being the Huge Large Quasar Group (Huge-LQG; \cite{clo12}) 
having a length of more than 1400 Mpc. 

As large as it appears to be, the Her-CrB GW does not necessarily
violate the basic assumptions of the cosmological principle
(the assumptions of a homogeneous and isotropic universe).
Theoretical large-scale structure models indicate
that some structures will exceed the End of Greatness on
purely statistical grounds (\cite{nada13}), and this may be one such structure
(albeit a very large one).
Along these lines, this may not be a single structure, but 
overlapping smaller adjacent and/or line-of-sight structures;
the small number of bursts currently found in the cluster limits 
our ability to angularly resolve it.
In other words, this may become a semantic issue at some point,
since a cluster of smaller structures might still be a larger structure.

This research was supported by the Hungarian OTKA grant NN111016 and by NASA EPSCoR grant NNX13AD28A. 
Discussions with L.G. Bal\'azs are also acknowledged.

\end{document}